\documentclass[conference]{IEEEtran}

\usepackage{etex}
\usepackage{booktabs}
\usepackage[utf8]{inputenc}
\usepackage[ruled,vlined]{algorithm2e}
\usepackage[T1]{fontenc}
\usepackage{microtype}
\usepackage{graphicx}
\usepackage{paralist}
\usepackage{tabularx}
\usepackage{balance}
\usepackage{multicol}
\usepackage{multirow}
\usepackage{pbox}
\usepackage{enumitem}
\usepackage{amssymb}
\usepackage{colortbl}
\usepackage{pifont}
\usepackage{xspace}
\usepackage{url}
\usepackage{tikz}
\usepackage{rotating}
\usepackage{float}
\usepackage{listings}
\usepackage[TABBOTCAP]{subfigure}
\usepackage{xcolor}
\usepackage{fontawesome}

\definecolor{gray}{rgb}{0.83,0.83,0.83}
\definecolor{green}{rgb}{0.56,0.93,0.56}

\newcommand{\ie}{\emph{i.e.,}\xspace}
\newcommand{\eg}{\emph{e.g.,}\xspace}
\newcommand{\etc}{etc.\xspace}
\newcommand{\etal}{\emph{et~al.}\xspace}
\newcommand{\secref}[1]{Section~\ref{#1}\xspace}

\newcommand{\figref}[1]{Fig.~\ref{#1}\xspace}

\newcommand{\tabref}[1]{Table~\ref{#1}\xspace}

\newboolean{showcomments}

\setboolean{showcomments}{true}

\ifthenelse{\boolean{showcomments}}
{\newcommand{\nb}[2]{
		\fbox{\bfseries\sffamily\scriptsize#1}
		{\sf\small$\blacktriangleright$\textit{#2}$\blacktriangleleft$}
	}
}
{\newcommand{\nb}[2]{}
}

\definecolor{light-gray}{gray}{0.92}

\makeatletter
\begin{document}

\title{On Learning Meaningful Code Changes via\\Neural Machine Translation\vspace{0cm}}

\author{
	\IEEEauthorblockN{Michele~Tufano\IEEEauthorrefmark{1},~Jevgenija~Pantiuchina\IEEEauthorrefmark{2},~Cody~Watson\IEEEauthorrefmark{1},~Gabriele~Bavota\IEEEauthorrefmark{2},~Denys~Poshyvanyk\IEEEauthorrefmark{1}}
	\IEEEauthorblockA{\IEEEauthorrefmark{1}College~of~William~and~Mary,~Williamsburg,~Virginia,~USA\\Email:~\{mtufano,~cawatson,~denys\}@cs.wm.edu}
	\IEEEauthorblockA{\IEEEauthorrefmark{2}Universit\`{a} della Svizzera italiana (USI),~Lugano,~Switzerland\\Email:~\{gabriele.bavota,~jevgenija.pantiuchina\}@usi.ch}
}

\maketitle
\thispagestyle{plain}

\begin{abstract}
Recent years have seen the rise of Deep Learning (DL) techniques applied to source code. Researchers have exploited DL to automate several development and maintenance tasks, such as writing commit messages, generating comments and detecting vulnerabilities among others. One of the long lasting dreams of applying DL to source code is the possibility to automate non-trivial coding activities. While some steps in this direction have been taken (\eg learning how to fix bugs), there is still a glaring lack of empirical evidence on the types of code changes that can be learned and automatically applied by DL.

Our goal is to make this first important step by quantitatively and qualitatively investigating the ability of a Neural Machine Translation (NMT) model to learn how to automatically apply code changes implemented by developers during pull requests. We train and experiment with the NMT model on a set of 236k pairs of code components before and after the implementation of the changes provided in the pull requests. We show that, when applied in a narrow enough context (\ie small/medium-sized pairs of methods before/after the pull request changes), NMT can automatically replicate the changes implemented by developers during pull requests in up to 36\% of the cases. Moreover, our qualitative analysis shows that the model is capable of learning and replicating a wide variety of meaningful code changes, especially refactorings and bug-fixing activities. Our results pave the way for novel research in the area of DL on code, such as the automatic learning and applications of refactoring.
\end{abstract}


\section{Introduction{\label{sec:introduction}}}
Several works recently focused on the use of advanced machine learning techniques on source code with the goal of (semi)automating several non-trivial tasks, including code completion \cite{White:2015:TDL:2820518.2820559}, generation of commit messages \cite{8115626}, method names \cite{Allamanis:2015:SAM}, code comments \cite{6693113}, defect prediction \cite{Wang:2016:ALS:2884781.2884804}, bug localization \cite{7372035} and fixing \cite{Tufano:2018:EIL:3238147.3240732}, clone detection \cite{White:2016:DLC:2970276.2970326}, code search \cite{Gu:2018:DCS:3180155.3180167}, and learning API templates \cite{Gu:2016:DAL:2950290.2950334}.

The rise of this research thread in the software engineering (SE) community is due to a combination of factors. The first is the vast availability of data, specifically source code, and its surrounding artifacts in open-source repositories. For instance, at the time of writing this paper, GitHub alone hosted 100M repositories, with over 200M merged pull requests (PRs) and 2B commits. Second, DL has become a useful tool due to its ability to learn categorization of data through the hidden layer architecture, making it especially proficient in feature detection \cite{deep-learning-article}. Specifically, Neural Machine Translation (NMT) has become a premier method for the translation of different languages, surpassing that of human interpretation \cite{DBLP:journals/corr/WuSCLNMKCGMKSJL16}. A similar principle applies to ``translating'' one piece of source code into another. Here, the ambiguity of translating makes this method extremely versatile: One can learn to translate buggy code into fixed code, English into Spanish, Java into C, \etc The third is the availability of (relatively) cheap hardware able to efficiently run DL infrastructures. %DL has made great strides in both ability and versatility when applied to SE tasks.\DENYS{do we really need this last sentence?} 

Despite all the work, only a few approaches have been proposed to automate non-trivial coding activities. Tufano \etal \cite{Tufano:2018:EIL:3238147.3240732} showed that DL can be used to automate bug-fixing activities. However, there is still a lack of empirical evidence about the types of code changes that can actually be learned and automatically applied by using DL. While most of the works applying DL in SE focus on quantitatively evaluating the performance of the devised technique (\eg How many bugs is our approach able to fix?), little qualitative analysis has been done to deeply investigate the meaningfulness of the output produced by DL-based approaches. In this paper, we make the first empirical step towards extensively investigating the ability of an NMT model to learn how to automatically apply code changes just as developers do in PRs. We harness NMT to automatically ``translate'' a code component from its state \emph{before} the implementation of the PR and \emph{after} the PR has been merged, thereby, emulating the combination of code changes that would be implemented by developers in PRs. 

We mine three large Gerrit \cite{Gerrit} code review repositories, namely Android \cite{gerrit-android}, Google Source \cite{gerrit-google}, and Ovirt \cite{gerrit-ovirt}. In total, these repositories host code reviews related to 339 sub-projects. We collected from these projects 78,981 merged PRs that underwent code review. We only considered merged and reviewed PRs for three reasons. First, we wanted to ensure that an NMT model is learning \textit{meaningful changes}, thus, justifying the choice of mining ``reviewed PRs'' as opposed to any change committed in the versioning system. Second, given the deep qualitative focus of our study, we wanted to analyze the discussions carried out in the code review process to better understand the types of changes learned by our approach. Indeed, while for commits we would only have commit notes accompanying them, with a reviewed PR we can count on a rich qualitative data explaining the rationale behind the implemented changes. Third, we only focus on merged PRs, since the code before and after (\ie merged) the PR is available. This is not the case for abandoned PRs. We extract method-level AST edit operations from these PRs using fine-grained source code differencing \cite{Falleri:ase2014}. This resulted in 239,522 method pairs, each of them representing the method before (PR not submitted) and after (PR merged) the PR process. An Encoder-Decoder Recurrent Neural Network (RNN) is then used to learn the code transformations performed by developers during PR activities.

We demonstrate a quantitative and qualitative evaluation of the NMT model. For the quantitative analysis, we assessed its ability in modifying the project's code exactly as done by developers during real PRs. This means that we compare, for the same code components, the output of the manually implemented changes and of the output of the NMT model. The qualitative analysis aims instead at distilling a taxonomy of meaningful code transformations that the model was able to automatically learn from the training data --- see \figref{fig:taxonomy}.

The achieved results indicate that, in its best configuration, the NMT model is able to inject the same code transformations that are implemented by developers in PRs in 16-36\% of cases, depending on the number of possible solutions that it is required to produce using beam search \cite{Raychev:2014:CCS:2594291.2594321}. Moreover, the extracted taxonomy shows that the model is able to learn a rich variety of meaningful code transformations, automatically fixing bugs and refactoring code as humans would do. As explained in \secref{sec:design}, these results have been achieved in a quite narrow context (\ie we only considered pairs of small/medium methods before/after the implementation of the changes carried by the PR), and this is also one of the reasons why our infrastructure mostly learned bug-fixing and refactoring activities (as opposed to the implementation of new features). However, we believe that our results clearly show the potential of NMT for learning and automating non-trivial code changes and therefore can pave the way to more research targeting the automation of code changes (\eg approaches designed to learn and apply refactorings). To foster research in this direction, we make publicly available the complete datasets, source code, tools, and raw data used in our experiments \cite{replication}.

\section{Approach{\label{sec:approach}}}
Our approach starts with mining PRs from three large Gerrit repositories (Sec. \ref{sec:mining}). We extract the source code \textit{before} and \textit{after} the PRs are merged. We pair pre-PR and post-PR methods, where each pair serves as an example of a meaningful change (Sec. \ref{sec:extraction}). Method pairs are then abstracted, filtered, and organized in datasets (Sec. \ref{sec:abstraction}). We train our model to \textit{translate} the version of the code \textit{before} the PR into the one \textit{after} the PR, to emulate the code change (Sec. \ref{sec:learning}). Finally, NMT's output model is concretized into real code (Sec. \ref{sec:concretization}).

\subsection{Code Reviews Mining}
\label{sec:mining}
We built a Gerrit crawler to collect the PR data needed to train the NMT model. Given a Gerrit server, the crawler extracts the list of projects hosted on it. Then, for each project, the crawler retrieves the list of all PRs submitted for review and having ``merged'' as the final status. We then process each merged PR $P$ using the following steps. First, let us define the set of Java files submitted in $P$ as $F_S = \{F_1, F_2, \dots, F_n\}$. We ignore non-Java files, since our NMT model only supports Java. For each file in $F_S$, we use the Gerrit API to retrieve their version before the changes implemented in the PR. The crawler discards new files created in the PR (\ie not existing before the PR) since we cannot learn any code transformation from them (we need the code before/after the PR to learn changes implemented by developers). Then, Gerrit API is used to retrieve the merged file versions impacted by the PR. The two (before/after) file sets might not be exactly the same, due to files created/deleted during the review process. 

The output of the crawler is, for each PR, the version of the files impacted before (pre-PR) and after (post-PR, merged) the PR. At the end of the mining process we obtain three datasets of PRs: $PR_{Ovirt}$, $PR_{Android}$, and $PR_{Google}$.

\subsection{Code Extraction}
\label{sec:extraction}

Each mined PR is represented as $pr = \{(f_1, \dots, f_n),(f^{'}_1, \dots, f^{'}_m) \}$, where $f_1, \dots, f_n$ are the source code files \textit{before} the PR, and $f^{'}_1, \dots, f^{'}_m$ are code files \textit{after} the PR. As previously explained, the two sets may or may not be the same size, since files could have been added or removed during the PR process. In the first step, we rely on GumTreeDiff \cite{Falleri:ase2014} to establish the file-to-file mapping, performed using semantic anchors, between pre- and post-PR files and disregarding any file added/removed during the code review process. After this step, each PR is stored in the format $pr = \{(f_1, \dots, f_k),(f^{'}_1, \dots, f^{'}_k) \}$, where $f_i$ is the file before and $f^{'}_i$ the corresponding version of the file after the PR. Next, each pair of files $(f_i, f^{'}_i)$ is again analyzed using GumTreeDiff, which establishes method-to-method mapping and identifies AST operations performed between two versions of the same method. We select only pairs of methods for which the code after the PR has been changed with respect to the code before the PR. Then, each PR is represented as a list of paired methods $pr = \{(m_b, m_a)_1, \dots, (m_b, m_a)_n \}$, where each pair $(m_b, m_a)_i$ contains the method \textit{before} the PR ($m_b$) and the method \textit{after} the PR ($m_a$). These are examples of changes used to train an NMT model to translate $m_b$ in $m_a$.

We use the method-level granularity for several reasons: (i) methods implement a single functionality and provide enough context for a meaningful code transformation; (ii) file-level code changes are still possible by composing multiple method-level code transformations; (iii) files represent large corpus of text, with potentially many lines of untouched code during the PR, which would hinder our goal to train a NMT model.

In this paper we only study code changes which modify existing methods, disregarding code changes that involve the creation or deletion of entire methods/files (see \secref{sec:threats}).

\subsection{Code Abstraction \& Filtering}
\label{sec:abstraction}

NMT models generate sequences of tokens by computing probability distributions over words. They can become very slow or imprecise when dealing with a large vocabulary comprised of many possible output tokens. This problem has been addressed by artificially limiting the vocabulary size, considering only most common words, assigning special tokens (\eg \texttt{UNK}) to rare words or by learning subword units and splitting the words into constituent tokens \cite{DBLP:journals/corr/MiWI16, DBLP:journals/corr/WuSCLNMKCGMKSJL16}.

The problem of large vocabularies (a.k.a. open vocabulary) is well known in the Natural Language Processing (NLP) field, where languages such as English or Chinese can have hundreds of thousands of words. This problem is even more pronounced for source code. As a matter of fact, developers are not limited to a finite dictionary of words to represent source code, rather, they can generate a potentially infinite amount of novel identifiers and literals. Table \ref{tab:vocab} shows the number of unique tokens identified in the source code of the three datasets. The vocabulary of the datasets ranges between 42k and 267k, while the combined vocabulary of the three datasets exceeds 370k unique tokens. In comparison, the Oxford English Dictionary contains entries for 171,476 words \cite{oxford-dict}.

\begin{table}[t]
	\scriptsize
	\caption{Vocabularies}
	\vspace{-0.3cm}
	\label{tab:vocab}
	\centering
	\begin{tabular}{lrr}
		\toprule
		Dataset & Vocabulary & Abstracted Vocabulary\\
		\midrule
		Google & 42,430  & 373 \\
		Android & 266,663 & 429 \\
		Ovirt  & 81,627 & 351 \\
		\midrule
		All  & 370,519 & 740\\
		\bottomrule
	\end{tabular}
	\vspace{-0.5cm}
\end{table}

In order to allow the training of an NMT model, we need a way to reduce the vocabulary while still retaining semantic information of the source code. We employ an abstraction process which relies on the following observations regarding code changes: (i) several chunks of code might remain untouched; (ii) developers tend to reuse identifiers and literals already present in the code; (iii) frequent identifiers (\ie common API calls and variable names) and literals (\eg 0, 1, ``foo'') are likely to be introduced in code changes. 

We start by computing the top-300 most frequent identifiers (\ie type, method, and variable names) and literals (\ie int, double, char, string values) used in the source code for each of the three datasets. This set contains frequent types, API calls, variable names and common literal values (\eg \texttt{0, 1, "\textbackslash n"}) that we want to keep in our vocabulary.

Subsequently, we abstract the source code of the method pairs by means of a process that replaces identifiers and literals with reusable IDs. The source code of a method is fed to a lexer, built on top of ANTLR \cite{Parr:2013}, which tokenizes the raw code into a stream of tokens. This stream of tokens is then fed into a Java parser, which discerns the role of each identifier (\ie whether it represents a variable, method, or type name) and the type of a literal. Each unique identifier and literal is mapped to an ID, having the form of \texttt{CATEGORY\_\#}, where \texttt{CATEGORY} represents the type of identifier or literal (\ie \texttt{TYPE}, \texttt{METHOD}, \texttt{VAR}, \texttt{INT}, \texttt{FLOAT}, \texttt{CHAR}, \texttt{STRING}) and \# is a numerical ID generated sequentially for each unique type of instance within that category (\eg the first method will receive \texttt{METHOD\_0}, the third integer value \texttt{INT\_2}, \etc). These IDs are used in place of identifiers and literals in the abstracted code, while the mapping between IDs and actual identifier/literal values is saved in a map $M$, which allows us to map back the IDs in the code concretization phase (\secref{sec:concretization}). During the abstraction process, we replace all identifiers/literals with IDs, except for the list of 300 most frequent identifiers and literals, for which we keep the original token value in the corpus.

Given a method pair $(m_b, m_a)$, the method $m_b$ is abstracted first. Then, using the same mapping $M$ generated during the abstraction of $m_b$, the method $m_a$ is abstracted in such a way that identifiers/literals already available in $M$ will use the same ID, while new identifiers/literals introduced in $m_a$ (and not available in $m_b$) will receive a new ID. At the end of this process, from the original method pair $(m_b, m_a)$ we obtain the abstracted method pair $(am_b, am_a)$. 

We allow IDs to be reused across different method pairs (\eg the first method name will always receive the ID \texttt{METHOD\_0}), therefore leading to an overall reduction of the vocabulary size. The third column of Table \ref{tab:vocab} reports the vocabulary size after the abstraction process, which shows a significant reduction in the number of unique tokens in the corpus.
In particular, after the abstraction process, the vocabulary contains: (i) Java keywords; (ii) top-300 identifiers/literals; (iii) reusable IDs. It is worth noting that the last row in Table \ref{tab:vocab} (\ie All) does not represent the cumulative sum, but rather the count of unique tokens when the three dataset corpora are merged.

Having a relatively small vocabulary allows the NMT model to focus on learning patterns of code transformations that are common in different contexts. Moreover, the use of frequent identifiers and literals allows the NMT model to learn typical changes (\eg \texttt{if(i>1)} to \texttt{if(i>0)}) and introduce API calls based on other API calls already available in the code.

After the abstraction process, we filter out method pairs from which the NMT model would not be able to learn code transformations that will result in actual source code. To understand the reasoning behind this filtering, it is important to understand the real use case scenarios.  When the NMT model receives the source code of the method $am_b$, it can only perform code transformations that involve: (i) Java keywords; (ii) frequent identifiers/literals; (iii) identifiers and literals already available in $m_b$. Therefore, we disregard method pairs where $m_a$ contains tokens not listed in the three aforementioned categories, since the model would have to synthesize new identifies or literals not previously seen.

In the future, we plan to increase the number of frequent identifiers and literals used in the vocabulary with the aim of learning code transformations from as many method pairs as possible. We also filter out those method pairs such that $am_b = am_a$, meaning the abstracted code before and after the PR appear the same. We remove these instances since the NMT model would not learn any code transformation.

Next, we partition the method pairs in small and medium pairs, based on their size measured in the number of tokens. In particular, small method pairs are those no longer than 50 tokens, while we consider medium pairs those having a length between 50-100 tokens. In this stage, we disregard longer method pairs. We discuss this limitation in \secref{sec:threats}.

Table \ref{tab:datasets} shows the number of method pairs, after the abstraction and filtering process, for each dataset and the combined one (\ie All). Each of the four datasets is then randomly partitioned into training (80\%), validation (10\%), and test (10\%) sets. Before doing so, we make sure to remove any duplicate method pairs, to ensure that none of the method pairs in the test set have been seen during the training phase.

\begin{table}[t]
	\scriptsize
	\caption{Datasets
	}
	\vspace{-0.3cm}
	\label{tab:datasets}
	\centering
	\begin{tabular}{lrr}
		\toprule
		Dataset & $M_{small}$ & $M_{medium}$\\
		\midrule
		Google & 2,165 & 2,286\\
		Android & 4,162 & 3,617\\
		Ovirt & 4,456 & 5,088\\
		\midrule
		All & 10,783 & 10,991\\
		\bottomrule
	\end{tabular}
	\vspace{-0.4cm}
\end{table}

\subsection{Learning Code Transformations}
\label{sec:learning}
In this section, we describe the NMT models we use to learn code transformations. In particular, we train these models to translate the abstracted code $am_b$ in $am_a$, effectively simulating the code change performed in the PR by developers.

\subsubsection{RNN Encoder-Decoder}
To build such models, we rely on an RNN Encoder-Decoder architecture with attention mechanism \cite{DBLP:journals/corr/BahdanauCB14, DBLP:journals/corr/LuongPM15, DBLP:journals/corr/BritzGLL17}, commonly adopted in NMT tasks \cite{kalchbrenner-blunsom:2013:EMNLP, DBLP:journals/corr/SutskeverVL14, DBLP:journals/corr/ChoMGBSB14}. As the name suggests, this model consists of two major components: an RNN Encoder, which \textit{encodes} a sequence of tokens \boldmath{$x$} into a vector representation, and an RNN Decoder, which \textit{decodes} the representation into another sequence of tokens $y$. During training, the model learns a conditional distribution over a (output) sequence conditioned on another (input) sequence of terms: \unboldmath$P(y_1,.., y_m | x_1,.., x_n)$, where the lengths $n$ and $m$ may differ. In our setting, given the sequence representing the abstract code before the PR $\mathbf{x} = am_b = (x_1,.., x_n)$ and a corresponding target sequence representing the abstract code after the PR $\mathbf{y} = am_a = (y_1,.., y_m)$, the model is trained to learn the conditional distribution:  $P(am_a | am_b) = P(y_1,.., y_m | x_1,.., x_n)$, where $x_i$ and $y_j$ are abstracted source tokens: Java keywords, separators, IDs, and frequent identifiers and literals. The Encoder takes as input a sequence $\mathbf{x} = (x_1,.., x_n)$ and produces a sequence of states $\mathbf{h} = (h_1,.., h_n)$. In particular, we adopt a bi-directional RNN Encoder \cite{DBLP:journals/corr/BahdanauCB14}, which is formed by a backward and a forward RNN. The RNNs process the sentence both from left-to-right and right-to-left, and are able to create sentence representations taking into account both past and future inputs \cite{DBLP:journals/corr/BritzGLL17}. The RNN Decoder predicts the probability of a target sequence $\mathbf{y} = (y_1,.., y_m)$ given $\mathbf{h}$. Specifically, the probability of each output token $y_i$ is computed based on: (i) the recurrent state $s_i$ in the Decoder; (ii) the previous $i-1$ tokens $(y_1,.., y_{i-1})$; and (iii) a context vector $c_i$. This vector $c_i$, also called attention vector, is computed as a weighted average of the states in $\mathbf{h}$: $c_i = \sum_{t=1}^{n}{a_{it}h_t}$ where the weights $a_{it}$ allow the model to pay more \textit{attention} to different parts of the input sequence, when predicting the token $y_i$. Encoder and Decoder are trained jointly by minimizing the negative log likelihood of the target tokens, using stochastic gradient descent.

\subsubsection{Beam Search Decoding}
For each method pair $(am_b, am_a)$ the model is trained to translate $am_b$ solely into the corresponding $am_a$. However, during testing, we would like to obtain \textit{multiple} possible translations. Precisely, given a piece of source code $m$ as input to the model, we would like to obtain $k$ possible translations of $m$. To this aim, we employ a decoding strategy called a Beam Search used in previous applications of DL \cite{Raychev:2014:CCS:2594291.2594321}. The major intuition behind a Beam Search decoding is that rather than predicting at each time step the token with the best probability, the decoding process keeps track of $k$ hypotheses (with $k$ being the beam size). Formally, let $\mathcal{H}_t$ be the set of $k$ hypotheses decoded until time step $t$:
$$
\mathcal{H}_t = \{ (\tilde{y}^{1}_1, \dots , \tilde{y}^{1}_t) , (\tilde{y}^{2}_1, \dots , \tilde{y}^{2}_t) , \dots , (\tilde{y}^{k}_1, \dots , \tilde{y}^{k}_t)   \}
$$
At the next time step $t+1$, for each hypothesis there will be $|V|$ possible $y_{t+1}$ terms ($V$ being the vocabulary), for a total of $k \cdot |V|$ possible hypotheses:
$$
\mathcal{C}_{t+1} = \bigcup\limits_{i=1}^{k}  \{ (\tilde{y}^{i}_1, \dots , \tilde{y}^{i}_t, v_1) , \dots , (\tilde{y}^{i}_1, \dots , \tilde{y}^{i}_t, v_{|V|})   \}
$$
From these candidate sets, the decoding process keeps the $k$ sequences with the highest probability. The process continues until each hypothesis reaches the special token representing the end of a sequence. We consider these $k$ final sentences as candidate patches for the buggy code.

\subsubsection{Hyperparameter Search}
\label{sec:hyperpar}
We tested ten configurations of the encoder-decoder architecture with different combinations of RNN Cells (LSTM \cite{Hochreiter:1997:LSM:1246443.1246450} and GRU \cite{DBLP:journals/corr/ChoMGBSB14}), number of layers (1, 2, 4) and units (256, 512) for the encoder/decoder, and the embedding size (256, 512). Bucketing and padding was used to deal with the variable length of the sequences. We trained the models for a maximum of 60k epochs, and selected the model's checkpoint before over-fitting the training data. To guide the selection of the best configuration, we used the loss function computed on the \textit{validation} set (not on the test set), while the results  are computed on the \textit{test} set.

\subsection{Code Concretization}
\label{sec:concretization}
In this final phase, the abstracted code generated as output by the NMT model is concretized by mapping back all the identifiers and literal IDs to their actual values. The process simply replaces each ID found in the abstracted code to the real identifier/literal associated with the ID and saved in the mapping $M$, for each method pair. The code is automatically indented and additional code style rules can be enforced during this stage. While we do not deal with comments, they could be reintroduced in this stage as well.

\section{Study Design{\label{sec:design}}}
The {\em goal} of this study is to empirically assess whether NMT can be used to learn a diverse and meaningful set of code changes. The {\em context} consists of a dataset of PRs and aims at answering two research questions (RQs). 

\subsection{RQ1: Can Neural Machine Translation be employed to learn meaningful code changes?}
We aim to empirically assess whether NMT is a viable approach to learn transformations of the code, as performed by developers in PRs. To this end, we use the eight datasets of method pairs listed in \tabref{tab:datasets}. Given a dataset, we train different configurations of the Encoder-Decoder models on the training set, then use the validation set to select the best performing configuration of the model. We then evaluate the validity of the model with the unseen instances of the test set. In total, we experiment with eight different models, one for each dataset in \tabref{tab:datasets} (\ie one model trained, configured, and evaluated on the Google dataset of small methods, one on the Google dataset of medium methods, \etc).

The evaluation is performed by the following methodology. Let $M$ be a trained model and $T$ be the test set of dataset $D$, we evaluate the model $M$ for each $(am_b, am_a) \in T$. Specifically, we feed the pre-PR abstract code $am_b$ to the model $M$, performing inference with Beam Search Decoding for a given beam size $k$. The model will generate $k$ different potential code transformations $CT = \{ct^1, \dots , ct^k\}$. We say that the model successfully predicted a code transformation if there exists a $ct^i \in CT$ such that $ct^i = am_a$ (\ie the abstract code generated by developers after the merging of the PR). We report the raw count and percentage of successfully predicted code changes in the test set, with $k= 1, 5, 10$. In other words, given a source code method that the model has never seen before, we evaluate the model's ability to correctly predict the code transformation that a developer performed by allowing the model to generate its best guess (\ie $k=1$) or the top-5 and top-10 best guesses.
It should be noted that while we count only perfect predictions, there are many other (slightly different) transformations that can still be viable and useful for developers. However, we discount these less-than-perfect predictions since it is not possible to automatically categorize those as viable and non-viable.

\subsection{RQ2: What types of meaningful code changes can be performed by the model?}
In this RQ we aim to qualitatively assess the types of code changes that the NMT model is able to generate. To this goal, we focus only on the successfully predicted code transformations generated by the model trained on the $All$ dataset, considering both small and medium sized methods. 

One of the authors manually investigated all the successfully predicted code transformations and described the code changes. Subsequently, a second author discussed and validated the described code changes. Finally, the five authors together defined -- and iteratively refined -- a taxonomy of code transformations successfully performed by the NMT model.

\section{Study Results{\label{sec:results}}}
\begin{figure*}
 \includegraphics[width=\linewidth]{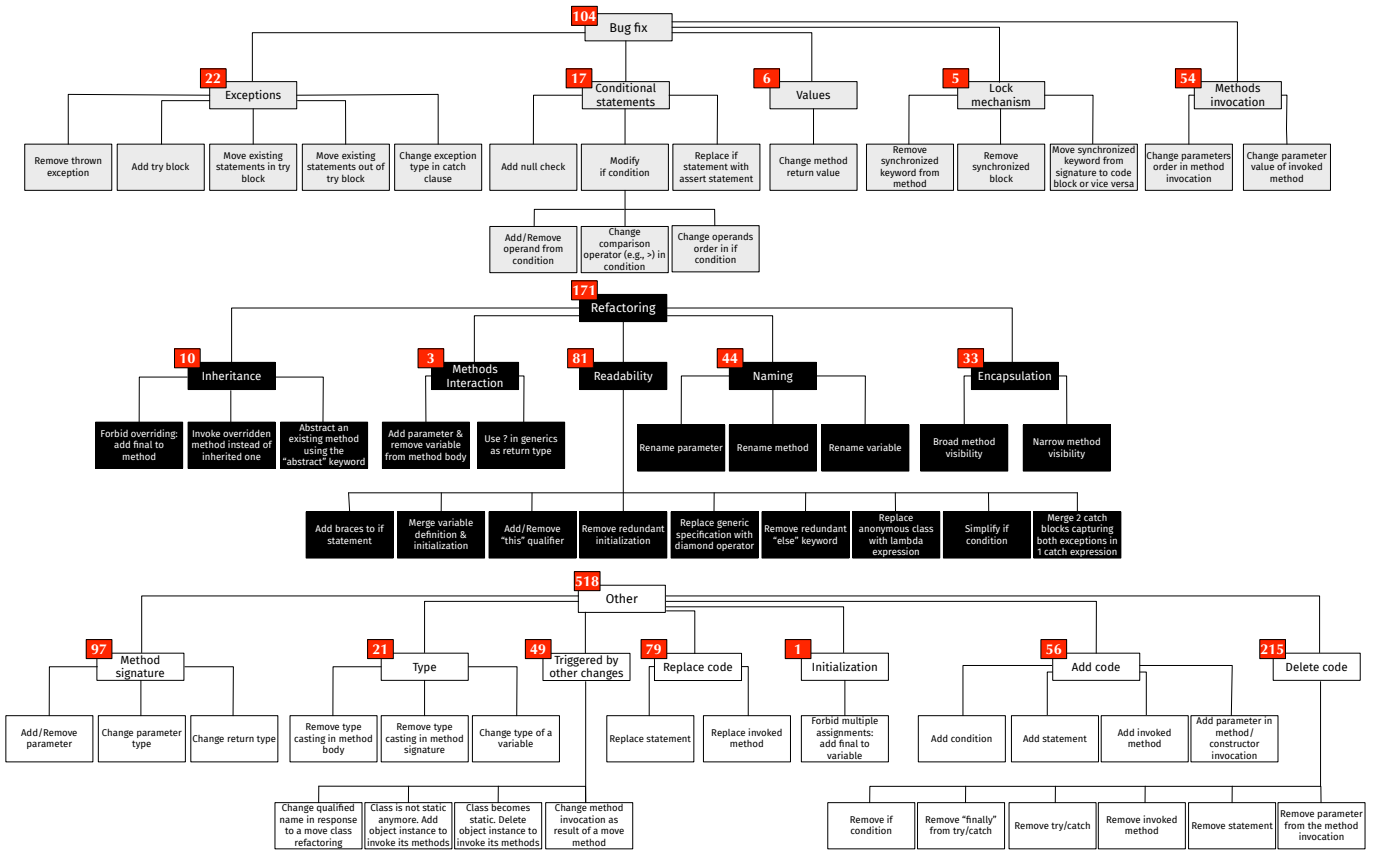}
 \caption{Taxonomy of code transformations learned by the NMT model}
 \label{fig:taxonomy}
 \vspace{-0.4cm}
\end{figure*}

\begin{table}[t]
	\scriptsize	
	\caption{Perfect Predictions}
	\vspace{-0.3cm}
	\label{tab:rq1}
	\centering
	\begin{tabular}{lccc}
		\toprule
		Dataset & Beam & $M_{small}$ & $M_{medium}$\\
		\midrule
		\multirow{3}{*}{Google} & 1 & 10 (4.62\%) & 7 (3.07\%)\\
		 & 5 & 17 (7.87\%) & 13 (5.70\%)\\
		 & 10 & 20 (9.25\%) & 17 (7.45\%)\\
		\midrule
		\multirow{3}{*}{Android} & 1 & 40 (9.61\%) & 51 (14.12\%)\\
		& 5 & 71 (17.06\%) & 73 (20.22\%)\\
		& 10 & 79 (18.99\%) & 76 (21.05\%)\\
		\midrule
		\multirow{3}{*}{Ovirt} & 1 & 55 (12.35\%) & 60 (11.78\%)\\
		& 5 & 93 (20.89\%) & 90 (17.68\%)\\
		& 10 & 113 (25.39\%) & 102 (20.03\%)\\
		\midrule
		\multirow{3}{*}{All} & 1 & 228 (21.16\%) & 178 (16.21\%)\\
		& 5 & 349 (32.40\%) & 306 (27.86\%)\\
		& 10 & 388 (36.02\%) & 334 (30.41\%)\\
		\bottomrule
	\end{tabular}
	 \vspace{-0.3cm}
\end{table}

\subsection{RQ1: Can Neural Machine Translation be employed to learn meaningful code changes?}

\tabref{tab:rq1} reports the perfect predictions (\ie successfully predicted code transformations) by the NMT models, in terms of raw numbers and percentages of the test sets. When we allow the models to generate only a single translation (\ie beam = 1), they are able to predict the same code transformation performed by the developers in 3\% up to 21\% of the cases. It is worth noting how the model trained on the combined datasets (\ie All) is able to outperform all the other single-dataset model, achieving impressive results even with a single guess (21.16\% for small and 16.21\% for medium methods). This result shows that NMT models are able to learn code transformations from a heterogeneous set of examples belonging to different datasets. Moreover, this also provides preliminary evidence that transfer learning would be possible for such models.

On the other end of the spectrum, the poor performance of the models trained on Google's dataset could be explained by the limited amount of training data (see \tabref{tab:datasets}) with respect to the other datasets.

When we allow the same models to generate multiple translations of the code (\ie 5 and 10), we observe a significant increase in perfect predictions across all models. On average, 1 out of 3 code transformations can be generated and perfectly predicted by the NMT model trained on the combined dataset. The model can generate 10 transformations in less than one second on a consumer-level GPU.\smallskip

{\em \underline{\textbf{Summary for RQ$_{1}$}.}} NMT models are able to learn meaningful code changes and perfectly predict code transformations in up to 21\% of the cases when only one translation is generated, and up to 36\% when 10 possible guesses are generated.

\subsection{RQ2: What types of meaningful code changes can be performed by the model?}

Here we focus on the 722 (388+334) perfect predictions generated by the model trained on the whole dataset, \ie All, with beam size equals 10. These perfect predictions were the results of 216 unique types of AST operations, as detected by GumTreeDiff, that the model was able to emulate. The complete list is available in our replication package \cite{replication}.

\figref{fig:taxonomy} shows the taxonomy of code transformations that we derived by manually analyzing the 722 perfect predictions. Note that a single perfect prediction can include multiple types of changes falling into different categories of our taxonomy (\eg a refactoring and a bug fix implemented in the same code transformation). For this reason, the sum of the classified changes in \figref{fig:taxonomy} is 793. The taxonomy is composed of three sub-trees, grouping code transformations related to bug fixing, refactoring, and ``other'' types of changes. The latter includes code transformations that the  model correctly performed (\ie those replicating what was actually done by developers during the reviewed PRs) but for which we were unable to understand the rationale behind the code transformation (\ie why it was performed). We preferred to adopt a conservative approach and categorize transformations into ``refactoring'' and ``bug-fix'' sub-trees only when we can confidently link to these types of activities. Also, for 27 transformations, the authors did not agree on the type of code change and, hence, we excluded them from our taxonomy (related to 695 perfect predictions).

Here, we qualitatively discuss interesting examples (indicated using the~\faCodeFork~icon) of code transformations belonging to our taxonomy. We do not report examples for all possible categories of changes learned by the model due to lack of space. Yet, the complete set of perfect predictions and their classification is available in our replication package \cite{replication}. 

\subsection{Refactoring}
We grouped in the refactoring sub-tree, all code transformations that modify the internal structure of the system by improving one or more of its non-functional attributes (\eg readability) without changing the system's external behavior. We categorized transformations into five sub-categories. \smallskip

\subsubsection{Inheritance} Refactorings that impact how the inheritance mechanism is used in the code. We found three types of refactorings related to inheritance: (i) forbid method overriding by adding the $\mathtt{final}$ keyword to the method declaration; (ii) invoke overriding method instead of overridden by removing the $\mathtt{super}$ keyword to the method invocation; and (iii) making a method abstract through the $\mathtt{abstract}$ keyword and deleting the method body.\smallskip

\noindent \faCodeFork~\textbf{Existing method declared as final \cite{android-example-final}}. In the $\mathtt{DirectByteBuffer}$ class of Android, the NMT model added to the signature of the $\mathtt{getLong(int)}$ method the $\mathtt{final}$ keyword. As stated by the developer implementing the PR: ``\emph{DirectByteBuffer cannot be final, but we can declare most methods final to make it easier to reason about}''. \smallskip

\noindent \faCodeFork~\textbf{Removed unnecessary ``super'' specifier \cite{ovirt-redundant-super}.} A PR in the Ovirt core subsystem was performed to clean up the class $\mathtt{RandomUtil}$, that extends Java class $\mathtt{java.util.Random}$. The $\mathtt{nextShort()}$ method implemented in the refactored class was invoking $\mathtt{nextInt()}$ of the base class through the use of the $\mathtt{super}$ java specifier. However, such a specifier was redundant because $\mathtt{nextInt()}$ was not overridden in $\mathtt{RandomUtil}$. Thus, it was removed by the developer: ``\emph{Using this modifier has no meaning in the context that was removed}''.\smallskip

\noindent \faCodeFork~\textbf{Existing method converted to abstract \cite{android-abstract-method}.}
	%https://android-review.googlesource.com/c/platform/libcore/+/675863
\begin{lstlisting}
float getFloatUnchecked(int index) {
  throw new UnsupportedOperationException();
}

abstract float getFloatUnchecked(int index);
\end{lstlisting}

\noindent The above code listing shows the code taken as input by the NMT model (top part, pre-PR) and produced as output (bottom, post-PR). The code transformation replicates the changes implemented by a developer in a PR, converting the $\mathtt{getFloatUnchecked}$ method into an abstract method, deleting its body. The rationale for this change is explained by the developer who implemented this change: The method $\mathtt{getFloatUnchecked}$ is overridden in all child classes of the abstract class implementing it and, thus, ``\emph{there is no need for the abstract base class to carry an implementation that throws $\mathtt{UnsupportedOperationException}$}''. The developer also mentions alternative solutions, such as moving this and similar methods into an interface, but concludes saying that the effort would be much higher. This case is interesting for at least two reasons. First, our model was able to learn a combination of code transformations needed to replicate the PR implemented by the developer (\ie add the $\mathtt{abstract}$ keyword \emph{and} delete the method body). Second, it shows the rich availability of information about the ``rationale'' for the implemented changes available in code review repositories. This could be exploited in the future to not only learn the code transformation, but also to justify it by automatically deriving the rationale from the developers' discussion.\smallskip

\subsubsection{Methods Interaction} These refactorings impact the way in which methods of the system interact, and include (i) \emph{add parameter} refactoring (\ie a value previously computed in the method body is now passed as parameter to it), and (ii) broadening the return type of a method by using the Java wildcard ($\mathtt{?}$) symbol.\smallskip

\noindent \faCodeFork~\textbf{Method returns a broader generic type \cite{google-generic-type}.}
		%https://gerrit-review.googlesource.com/c/gerrit/+/127039
		
\begin{lstlisting}
<I> RestModifyView<P,I> post(P parent) throws [...];
		
RestModifyView<P,?> post(P parent) throws [...];
\end{lstlisting}

\noindent The code listing shows a change implemented in a PR done on the ``Google'' Gerrit repository and correctly replicated by the NMT model. The $\mathtt{post}$ method declaration was refactored to return a broader type and improve the usage of generics. As explained by the developer, this also allows to avoid the `unchecked' warnings from the five implementations of the $\mathtt{post}$ method present in the system, thus simplifying the code.  
	
\smallskip
\subsubsection{Naming} This category groups refactorings related to the renaming of methods, parameters, and variables. This is usually done to improve the expressiveness of identifiers and to better adhere to the coding style guidelines. Indeed, good identifiers improve readability, understandability and maintainability of source code~\cite{Allamanis:2015:SAM,Einar:2009}. \smallskip

\noindent \faCodeFork~\textbf{Rename method \cite{ovirt-rename-method}.}
One example of correctly learned rename method, is the one fixing a typo from the $\mathtt{OnSucess}$ method in the Ovirt system \cite{ovirt-rename-method}. In this case, the developer (and the NMT model) both suggested to rename the method in $\mathtt{OnSuccess}$. \smallskip

\noindent \faCodeFork~\textbf{Rename parameter \cite{android-rename-parameter}.}
A second example of renaming, is the renamed parameter proposed for the $\mathtt{endTrace(JMethod~type)}$ method in a PR impacting the $\mathtt{AbstractTracerBrush}$ class in the Android repository \cite{android-rename-parameter}. The developer here renamed several parameters ``for clarity'' and, in this case, renamed the $\mathtt{type}$ parameter into $\mathtt{method}$, to make it more descriptive and better reflect its aim. \smallskip

\subsubsection{Encapsulation} We found refactorings aimed at broadening and narrowing the visibility of methods (see \figref{fig:taxonomy}). This can be done by modifying the access modifiers (\eg changing a public method to a private one). \smallskip

\noindent \faCodeFork~\textbf{Broadening \cite{android-broadening-method-visibility} and narrowing \cite{google-narrowing-method-visibility} method visibility.} An example of a method, for which our model recommended to broaden its visibility from $\mathtt{private}$ to $\mathtt{public}$, is the $\mathtt{of}$ method from the $\mathtt{Key}$ Android class \cite{android-broadening-method-visibility}. This change was done in a PR to allow the usage of the method from outside the class, since the developer needed it to implement a new feature. 

The visibility was instead narrowed from $\mathtt{public}$ to $\mathtt{private}$ in the context of a refactoring performed by a developer to make ``\emph{more methods private}'' \cite{google-narrowing-method-visibility}. This change impacted the $\mathtt{CurrentUser.getUser()}$ method from the Google repository, and the rationale for this change correctly replicated by the NMT model was that the $\mathtt{getUser()}$ method was only used in one location in the system outside of its class. However, in that location the value of ``\emph{the user is already known}'', thus do not really requiring the invocation of $\mathtt{getUser()}$. \smallskip

\subsubsection{Readability} Readable code is easier to understand and maintain \cite{Simone:2016:read1}. We found several types of code transformations learned by the model and targeting the improvement of code readability. This includes: (i) braces added to $\mathtt{if}$ statements with the only goal of clearly delimiting their scope; (ii) the merging of two statements defining (\eg $\mathtt{String}$ $\mathtt{address;}$) and initializing (\eg $\mathtt{address=getAddess();}$) a variable into a single statement doing both (\eg $\mathtt{String}$ $\mathtt{address=getAddess();}$); (iii) the addition/removal of the $\mathtt{this}$ qualifier, to match the project's coding standards; (iv) reducing the verbosity of a generic declaration by using the Java diamond operator (\eg $\mathtt{Map<String,}$ $\mathtt{List<String>>}$ $\mathtt{mapping=new}$ $\mathtt{HashMap<String,}$ $\mathtt{List<String>>()}$ becomes $\mathtt{Map<String,}$ $\mathtt{List<String>>}$ $\mathtt{mapping=new}$ $\mathtt{HashMap<>}$); (v) remove redundant $\mathtt{else}$ keywords from $\mathtt{if}$ statements (\ie when the code delimited by the $\mathtt{else}$ statement would be executed in any case); (vi) refactoring anonymous classes implementing one method to lambda expressions, to make the code more readable \cite{ovirt-anonymous-lambda}; (vii) simplifying boolean expressions (\eg $\mathtt{if(x == true)}$ becomes $\mathtt{if(x)}$, where $\mathtt{x}$ is a boolean variable); and (viii) merging two catch blocks capturing different exceptions into one catch block capturing both exceptions using the $\mathtt{or}$ operator \cite{android-merging-catches}.\smallskip

\noindent \faCodeFork~\textbf{Anonymous class replaced with lambda expression \cite{ovirt-anonymous-lambda}.} 
	%https://gerrit.ovirt.org/#/c/50859/	
	\begin{lstlisting}
public boolean isDiskExist([...]) {
  return execute(new java.util.concurrent.Callable<java.lang.Boolean>() {
    @java.lang.Override
    public java.lang.Boolean call() { try {[...]} } });	}
	
public boolean isDiskExist([...]) {
  return execute(() -> { try {[...]} }); }
	\end{lstlisting}
	
\noindent In the above code listing, the NMT model automatically replaces an anonymous class (top part, pre-PR) with a lambda expression (bottom part, post-PR), replicating changes made by Ovirt's developers during the transitions of the code through Java 8. The new syntax is more compact and readable. \smallskip

\newpage

\noindent \faCodeFork~\textbf{Merging catch blocks capturing different exceptions \cite{android-merging-catches}.} 
	%	https://android-review.googlesource.com/c/platform/libcore/+/244295/4/ojluni/src/main/java/java/lang/Integer.java
	\begin{lstlisting}
public static Integer getInteger(String nm, Integer val) {
  [...] 
  try {[...]}  
  catch (IllegalArgumentException e) { } 
  catch (NullPointerException e) { } 
}
	
public static Integer getInteger(String nm, Integer val) {
  [...] 
  try {[...]}  
  catch (IllegalArgumentException | NullPointerException e) { }
}
	
	\end{lstlisting}
	\noindent As part of a PR implementing several changes, the two $\mathtt{catch}$ blocks of the $\mathtt{getInteger}$ method were merged by the developer into a single $\mathtt{catch}$ block (see the code above). The NMT model was able to replicate such a code transformation that is only meaningful when an exception is caught and the resulting code that is executed is the same for both instances of the exception (as in this case). This code change, while simple from a developer's perspective, is not trivial to learn due to the several transformations to implement (\ie removal of the two $\mathtt{catch}$ blocks and implementation of a new $\mathtt{catch}$ block using the $\mathtt{|}$ or operator) and to the ``pre-condition'' to check (\ie the same behavior implemented in the catch blocks).
	
\subsection{Bug Fix}
	
Changes in the ``bug fix'' subtree (see \figref{fig:taxonomy}) include changes implemented with the goal of fixing a specific bug which has been introduced in the past. The learned code transformations are organized here into five sub-categories, grouping changes related to bug fixes that deal with (i) exception handling, (ii) the addition/modification of conditional statements, (iii) changes in the value returned by a method, (iv) the handling of lock mechanisms, and (v) wrong method invocations. \smallskip
	
\subsubsection{Exception} This category of changes is further specialized into several subcategories (see \figref{fig:taxonomy}) including (i) the addition/delation of thrown exceptions; (ii) the addition of $\mathtt{try-catch/finally}$ blocks \cite{android-add-catch}; (iii) narrowing or broadening the scope of the $\mathtt{try}$ block by moving the existing statements inside/outside the block \cite{android-narrow-catch}; (iv) changing the exception type in the catch clause to a narrower type (\eg replacing $\mathtt{Throwable}$ with $\mathtt{RuntimeException}$).\smallskip

\noindent \faCodeFork~\textbf{Add try-catch block \cite{android-add-catch}.} 
	% https://android-review.googlesource.com/c/platform/libcore/+/283122
	\begin{lstlisting}
public void test_getPort() throws IOException {
  DatagramSocket theSocket = new DatagramSocket();
  [...]
}

public void test_getPort() throws IOException {
  try (DatagramSocket theSocket = new DatagramSocket()) { 
  [...]
  }
}
	\end{lstlisting}
\noindent The above code from the Android repository, shows the change implemented in a PR aimed at fixing ``\emph{resource leakages in tests}''. The transformation performed by the NMT model wrapped the creation and usage of a $\mathtt{DatagramSocket}$ object into a $\mathtt{try-with-resources}$ block. This way $\mathtt{theSocket.close()}$ will be automatically invoked (or an exception will be thrown), thus avoiding resource leakage. \smallskip

\noindent \faCodeFork~\textbf{Narrowed the scope of try block \cite{android-narrow-catch}.} {
		%https://android-review.googlesource.com/c/platform/libcore/+/148551
		\begin{lstlisting}
public void testGet_NullPointerException() {
  try {
    ConcurrentHashMap c = new ConcurrentHashMap(5);
    c.get(null);
    shouldThrow();
  } catch (java.lang.NullPointerException success) {}
}
		
public void testGet_NullPointerException() {
  ConcurrentHashMap c = new ConcurrentHashMap(5);
  try {
    c.get(null);
    shouldThrow();
  } catch (java.lang.NullPointerException success) {}
}
		
	\end{lstlisting}
	
\noindent Another change replicated by the NMT model and impacting the Andorid test suite is the code transformation depicted above and moving the $\mathtt{ConcurrentHashMap}$ object instantiation outside of the $\mathtt{try}$ block. The reason for this change is the following. The involved test method is supposed to throw a $\mathtt{NullPointerException}$ in case $\mathtt{c.get(null)}$ is invoked. Yet, the test method would have also passed if the exception was thrown during the $\mathtt{c}$ instantiation. For this reason, the developer moved the object creation out of the $\mathtt{try}$ block. \smallskip

\subsubsection{Conditional statements} Several bugs can be fixed in conditional statements verifying that certain preconditions are met before specific actions are performed (\eg verifying that an object is not null before invoking one of its methods). \smallskip

\noindent \faCodeFork~\textbf{Added null check \cite{android-null-check}.} 
		%https://android-review.googlesource.com/c/platform/frameworks/base/+/382232
\begin{lstlisting}		
public void run() {
  mCallback.onConnectionStateChange(BluetoothGatt.this, GATT_FAILURE,
  BluetoothProfile.STATE_DISCONNECTED);
}
		
public void run() {
  if (mCallback != null) {
    mCallback.onConnectionStateChange(BluetoothGatt.this, GATT_FAILURE,
    BluetoothProfile.STATE_DISCONNECTED);
  }
}
\end{lstlisting}

\noindent The code listing shows the changes implemented in an Android PR to ``\emph{fix a NullPointerException when accessing mCallback in BluetoothGatt}''. The addition of the $\mathtt{if}$ statement implementing the null check allows the NMT model to fix the bug exactly as the developer did. \smallskip
	
\noindent \faCodeFork~\textbf{Change comparison operand \cite{android-if-comparison}.} 
	\begin{lstlisting}
public void reset(int i) {
  if ((i < 0) || (i >= mLen)) { [...] }
}

public void reset(int i) {
  if ((i < 0) || (i > mLen)) { [...] }
}
	\end{lstlisting}
\noindent A second example of a bug successfully fixed by the NMT model working on the conditional statements, impacted the API of the $\mathtt{FieldPacker}$ class. As explained by the developer, the PR contributed ``\emph{a fix to the FieldPacker.reset() API, which was not allowing the FieldPacker to ever point to the final entry in its buffer}''. This was done by changing the $>=$ operand to $>$ as shown in the code reported above.\smallskip
	
\subsubsection{Values} The only type of change we observed in this category is the change of methods' return value to fix a bug. This includes simple cases in which a $\mathtt{boolean}$ return value was changes from $\mathtt{false}$ to $\mathtt{true}$ (see \eg \cite{android-return-value}), as well as less obvious code transformations in which a constant return value was replaced with a field storing the current return value, \eg $\mathtt{return~"refs/my/config";}$ converted into $\mathtt{return~ref;}$, where $\mathtt{ref}$ is a variable initialized in the constructor \cite{google-return-value}.\smallskip

\subsubsection{Lock mechanism} These code changes are all related to the usage of the $\mathtt{synchronized}$ Java keyword in different parts of the code. These include its removal from a code block \cite{android-remove-synchronized}, from a method signature \cite{android-remove-synchronized-signature}, and moving the keyword from the method signature to a code block or \emph{vice versa} \cite{android-move-synch}. We do not discuss these transformations due to lack of space. \smallskip

\subsubsection{Methods invocation} These category groups code transformations fixing bugs by changing the order or value of parameters in method invocations. \smallskip
	
\noindent \faCodeFork~\textbf{Flipped parameters in assertEquals \cite{ovirt-flipped-parameters}.}
	%https://gerrit.ovirt.org/#/c/63570/
		\begin{lstlisting}
public void testConvertMBToBytes() {
  [...]
  org.junit.Assert.assertEquals(bytes, 3145728);
}
		
public void testConvertMBToBytes() {
  [...]
  org.junit.Assert.assertEquals(3145728, bytes);
\end{lstlisting}
	
\noindent In this example the developer fixed a bug in the test suite by flipping the order in which the parameters are passed to the $\mathtt{assertEquals}$ method.  In particular, while the assert method was expecting the pairs of parameters ($\mathtt{long~expected}$, $\mathtt{long~actual)}$,  test was passing the actual value first, thus invalidating the test. The fix, automatically applied by the NMT model, swaps the arguments of the $\mathtt{assertEquals}$. \smallskip
	
\subsection{Other}
As previously said, we assigned to the `Other' subtree those code transformations for which we were unable to clearly identify the motivation/reason. This subtree includes changes related to: (i) the method signature (added/removed/changed parameter or return type); (ii) types (removed type casting in method body or its signature, changed variable type); (iii) variable initialization; (iv) replaced statement/invoked method; (v) added code (condition, statement, invoked method, parameter); (vi) deleted code ($\mathtt{if}$ condition, $\mathtt{finally}$ block, $\mathtt{try-catch}$ block, invoked method, statement); (vii) changes triggered by the other changes (\eg static method call replaced with an instance method call or \emph{vice versa} --- see \figref{fig:taxonomy}). Note that, while we did not assign a specific ``meaning'' to these changes, due to a lack of domain knowledge of the involved systems, these are still perfect predictions that the NMT model performed. This means the code changes are identical to the ones implemented by developers in the PR. \smallskip

{\em \underline{\textbf{Summary for RQ$_{2}$}.}} Our results show the great potential of NMT for learning meaningful code changes. Indeed, the NMT model was able to learn and automatically apply a wide variety of code changes, mostly related to refactoring and bug-fixing activities. The fact that we did not find other types of changes, such as new feature implementation, might be due to the narrow context in which we applied our models (\ie methods of limited size), as well as to the fact that new features implemented in different classes and systems rarely exhibit recurring patterns (\ie recurring types of code changes) that the model can learn. More research is needed to make this further step ahead.

\section{Threats To Validity{\label{sec:threats}}}
\textbf{Construct validity.} We collected code components before and after pull requests through a crawler relying on the Gerrit API. The crawler has been extensively tested, and the manual analysis of the extracted pairs performed to define the taxonomy in \figref{fig:taxonomy} confirmed the correctness of the collected data.\smallskip

\textbf{Internal validity.} The performance of the NMT model might be influenced by the hyperparameter configuration we adopted. To ensure replicability, we explain in \secref{sec:approach} how hyperparameter search has been performed. 

We identified through the manual analysis the types of code transformations learned by the model. To mitigate subjectivity bias in such a process, the taxonomy definition has been done by one of the authors, double checked by a second author, and finally, the resulting taxonomy has been discussed among all authors to spot possible issues. Moreover, in case of doubts, the code transformation was categorized in the ``other'' subtree, in which we only observed the type of code change implemented, without conjecturing about the goal of the transformation. However, as in any manual process, errors are possible, and we cannot exclude the presence of misclassified code transformations in our taxonomy.\smallskip

\textbf{External validity.} We experimented with the NMT model on data related to Java programs only. However, the learning process is language-independent and the whole infrastructure can be instantiated for different programming languages by replacing the lexer, parser and AST differencing tools.

We only focused on methods having no more than 100 tokens. This is justified by the fact that we observe a higher density of method pairs with sizes less than 100 tokens in our dataset. The distribution also shows a long tail of large methods, which could be problematic when training a NMT model. Distribution and data can be accessed in our replication package \cite{replication}. Also, we only focus on learning code transformations of existing methods rather than the creation of new methods since these latter are (i) complex code changes that involve a higher level of understanding of the software system in its entirety; and (ii) not well-suited for NMT models since the translation would go from/to empty methods.

Finally, pull request data from three Gerrit repositories were used. While these repositories include hundreds of individual projects (thus ensuring a good external validity of our findings) our results might not generalize to other projects/languages.

\section{Related Work{\label{sec:related}}}
Deep Learning (DL) has recently become a useful tool to study different facets of software engineering. The unique representations allow for features to be discovered by the model rather than manual derivation. Due to the power of these representations, many works have applied these models to solve SE problems \cite{DBLP:journals/corr/GodefroidPS17}\cite{Alexandru:2016:GCS:2950290.2983951}\cite{8094414}\cite{7332513}\cite{8094426}\cite{8094415}\cite{8094428}\cite{Gupta2017DeepFixFC}. However, to the best of our knowledge, this is the first work that uses DL techniques to learn and create a taxonomy from a variety of code transformations taken from developers' PRs.

White \etal uses representation learning via a recursive autoencoder for the task of clone detection \cite{White:2016:DLC:2970276.2970326}. Each piece of code is represented as a stream of identifiers and literals, which they use as input to their DL model. Using a similar encoding, Tufano \etal encodes methods into four different representations, then the DL model evaluates how similar two pieces of code are based on their multiple representations \cite{Tufano:2018:DLS:3196398.3196431}. Another recent work by Tufano \etal applies NMT to bug-fixing patches the wild \cite{Tufano:2018:EIL:3238147.3240732}. This work applies a  similar approach, but rather than learning code transformations they attempt to learn bug-fixing commits to generate patches. These works are related to ours, since we use a similar code representation as input to the DL model, yet, we apply this methodology to learn as many code transformations as possible. 

White \etal also compare DL models with natural language processing models for the task of code suggestion. They show that DL models make code suggestions based upon contextual features learned by the model rather than the predictive power of the past \textit{n} tokens \cite{White:2015:TDL:2820518.2820559}. Further expanding upon the powerful, predictive capabilities of these models, Dam \etal presents DeepSoft, which is a DL-based architecture used for modeling software, code generation and software risk prediction \cite{Dam:2016:DVD:2950290.2983985}. 

DL has also been applied to the areas of bug triaging and localization. Lam \etal makes use of DL models and information retrieval to localize buggy files after a bug report is submitted. They use a revised Vector Space Model to create a representation the DL model can use to relate terms in a bug report to source code tokens \cite{7372035}. Likewise, to reduce the effort of bug triaging, Lee \etal applies a CNN to industrial software in order to properly triage bugs.  This approach uses word2vec to embed a summary and a description which the CNN then assigns to a developer
\cite{Lee:2017:ADL:3106237.3117776}. Related to software bugs, Wang \etal uses a Deep Belief Network (DBN) to learn semantic features from token vectors taken from a programs' ASTs. The network then predicts if the commit will be defective \cite{Wang:2016:ALS:2884781.2884804}.

Many DL usages aim to help developers with tasks outside of writing code. Choetkiertikul \etal proposes a DL architecture of long short-term memory and recurring highway network that aims to predict the effort estimation of a coding task \cite{8255666}. Another aid for developers is the ability to summarize a given segment of source code. To this point Allamanis \etal uses an Attentional Neural Network (ANN) with a convoluation layer in order to summarize pieces of source code into short, functional descriptions \cite{DBLP:journals/corr/AllamanisPS16}. Guo \etal develops a DL approach using RNNs and word embeddings to learn the sentence semantics of requirement artifacts, which helps to create traceability links in software projects \cite{Guo:2017:SES:3097368.3097370}. The last example of DL implementations that aid developers in the software development process is an approach developed by Gu \etal that helps to locate source code. This implementation uses NNs and natural language to embed code snippets with natural language descriptions into a high-dimensional vector space, helping developers locate source code based on natural language queries \cite{Gu:2018:DCS:3180155.3180167}.   

DL-based approaches have also been applied to more coding related tasks, one such task is accurate method and class naming.  Allamanis \etal uses a log-bilinear neural network to understand the context of a method or class and recommends a representative name that has not appeared in the training corpus \cite{Allamanis:2015:SAM:2786805.2786849}. Also helping with correct coding practices, Gu \etal uses an RNN encoder-decoder model to generate a series of correct API usages in source code based upon natural language queries. The learned semantics allow the model to associate natural language queries with a sequence of API usages \cite{Gu:2016:DAL:2950290.2950334}.

Recently we have seen DL infiltrate the mobile SE realm. Moran \etal uses a DL-based approach to automatically generate GUIs for mobile apps. In this approach, a deep CNN is used to help classify GUI components which can later be used to generate a mock GUI for a specific app \cite {DBLP:journals/corr/abs-1802-02312}.

Although DL approaches are prevalent in SE, this work is the first to apply DL to empirically evaluate the capability to learn code changes from developer PRs.
The previous work has shown that DL approaches can yield meaningful results given enough quality training data. Thus, we specifically apply NMT to automatically learn a variety of code transformations, from real pull requests, and create a meaningful taxonomy.

\section{Conclusion{\label{sec:conclusions}}}
We investigated the ability of NMT models to learn how to automatically apply code transformations. We first mine a dataset of complete and meaningful code changes performed by developers in merged pull requests, extracted from three Gerrit repositories. Then, we train NMT models to translate pre-PR code into post-PR code, effectively learning code transformations as performed by developers.

Our empirical analysis shows that NMT models are capable to learn code changes and perfectly predict code transformations in up to 21\% of the cases when only a single translation is generated, and up to 36\% when 10 possible guesses are generated. The results also highlight the ability of the models to learn from a heterogeneous set of PRs belonging to different datasets, indicating the possibility of transfer learning across projects and domains. The performed qualitative analysis also highlighted the ability of the NMT models to learn a wide variety of  code transformations, paving the way to further research in this field targeting the automatic learning and application of non-trivial code changes, such as refactoring operations. In that sense, we hope that the public availability of the source code of our infrastructure and of the data and tools we used \cite{replication}, can help in fostering research in this field.

\section{Acknowledgment}
\label{sec:acknowledgment}
% !TeX root = main.tex
This work is supported in part by the NSF CCF-1525902 and CCF-1815186 grants. Pantiuchina and Bavota thank the Swiss National Science foundation for the financial support through SNF Project JITRA, No. 172479.  Any opinions, findings, and conclusions expressed herein are the authors’ and do not necessarily reflect those of the sponsors.

\clearpage
\balance

\bibliographystyle{IEEEtranS}
\bibliography{main}

\end{document}